# Electric Current-Driven Microstructural Evolution in $SrTiO_3$

Jingjing Yang, Jian Luo *

Aiiso Yufeng Li Family Department of Chemical and Nano Engineering; Program in Materials Science and Engineering, University of California San Diego, La Jolla, California 92093, U.S.A.

## Abstract

Polycrystalline $SrTiO_3$ is employed as a model system to investigate microstructural evolution under applied electric currents. Under a substantial current density (e.g., ~40-41 mA/mm$^2$), well-aligned, elongated abnormal grains develop near the anode following a flash event, in contrast to previously reported cathode-side enhanced grain growth under negligible currents. The equivalent diameter of the abnormal grains increases linearly with time, deviating from classical parabolic grain growth kinetics. The applied current drives elemental redistribution near the anode, producing a Ti-rich region adjacent to a Sr-rich belt that migrates toward the cathode, from which the abnormal grains nucleate. Aberration-corrected scanning transmission electron microscopy and electron energy-loss spectroscopy reveal that the fast-moving grain boundaries (GBs) within the Ti-rich bulk region are Sr-enriched, O-depleted, and Ti-reduced ($Ti^{4+} \rightarrow Ti^{3+}$). An analysis based on the Brouwer diagram suggests the formation of *p–i–n* regions under the applied electric field. Conversion between electronic and ionic currents at the *p–i* and *i–n* junctions, field-driven precipitation and dissolution of the Sr-rich Ruddlesden-Popper phase, and field-driven migration of Sr and O vacancies collectively explain the elemental redistribution and redox-modulated migration of the Sr-rich belt. Incomplete redox reactions at the moving junctions create the moving Sr-rich belt and generate a locally reducing environment, consequently producing fast-moving, Sr-rich, reduced GBs. These findings reveal new mechanisms of electric current-driven defect-mediated microstructural evolution.

*Correspondence should be addressed to J.L. (email: jluo@ucsd.edu)

## Graphical Abstract

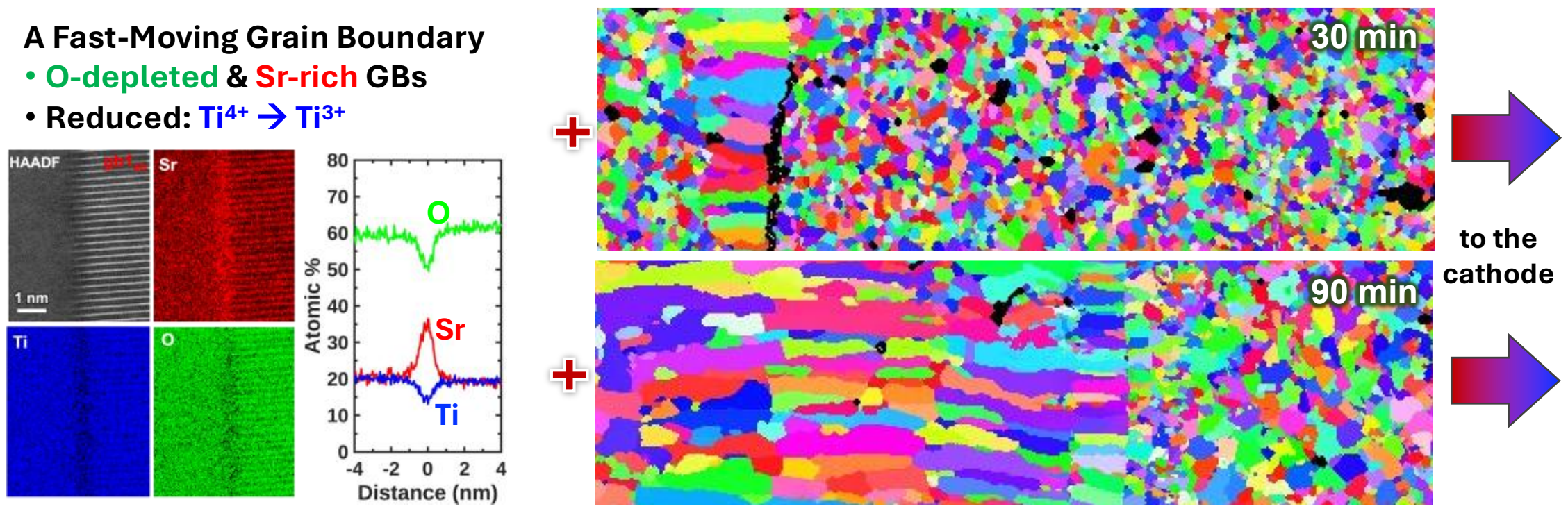

**Highlights**

- Well-aligned elongated grains grow near the anode in $SrTiO_3$ under currents.
- Field-defect interactions form an anode-side Ti-rich region and a Sr-rich belt.
- Redox reactions at *p*–*i* and *i*–*n* junctions drive Sr-rich belt migration.
- Local reduction creates fast-moving GBs with linear grain growth kinetics.
- Sr-rich, O-depleted, Ti-reduced GBs move rapidly in the Ti-rich bulk region.

# 1 Introduction

Electric fields and currents have been widely utilized in materials processing. For instance, flash sintering has attracted considerable interest for the rapid densification of various ceramics [1–9]. It is characterized by the onset of a flash event, marked by a sharp increase in power dissipation and a transition from voltage to current control. While follow-up studies have established that flash sintering is initiated by thermal runaway [9–12] and that ultrafast densification can occur without electric current flow through the specimen [10,12–16], electric fields and currents can alter microstructural evolution [6,9,17–31].

A growing body of work has demonstrated that electric fields and currents influence grain growth in a broad range of ionic materials, including NaCl [17], $Al_2O_3$ [32–34], yttria-stabilized zirconia (YSZ) [25,27,28,34–41], ZnO [42–45], and perovskite oxides [6,30,46–50]. However, the reported effects are highly diverse: grain growth can be either enhanced or suppressed, be preferentially promoted at either the anode or cathode, and exhibit steep spatial grain-size transitions, depending on the materials, dopants, and experimental conditions. In some studies, such effects have been attributed to field-induced changes in the local oxygen potential, which can modify bulk defects as well as the grain boundary (GB) segregation and space charges [28,38,46,48,51]. In addition, GB structural transitions induced by applied electric fields/currents have been reported [42,44,45,50]. Notably, applied electric fields have been used to create and control graded microstructures in ZnO [45] and $BaTiO_3$ [50] by inducing GB redox transitions through electrochemical coupling.

Systematic investigations of how applied electric fields and currents tailor microstructural evolution in perovskite oxides remain rare. Perovskite-structured strontium titanate ($SrTiO_3$), with wide applications in capacitors, varistors, field-effect transistors, solar cells, and photocatalysts [52–56], is an ideal model material owing to the extensive knowledge of its defect chemistry [57–60], GB structure and stoichiometry [61–63], space charges [61,64–66], and microstructure evolution [67–74]. To date, studies of electric-field effects on microstructural evolution in $SrTiO_3$ have reported enhanced grain growth exclusively at the cathode (the negative electrode), either under nominal electric-field strengths of 25–75 V/mm with current-blocking electrodes at 1350–1550 °C [46,51,75] or under a relatively low current density of ~4.5 mA/mm² during flash sintering at a sample temperature of 1200 °C [76]. Several mechanisms have been proposed. The accumulation of oxygen vacancies near the cathode is considered a dominant factor [46,51,75], consistent with the enhanced grain growth observed in reducing atmospheres [26,51,67,70–72,77,78]. Rheinheimer et al. further proposed that electromigration of oxygen vacancies toward the cathode reduces the space charge and solute drag at GBs, thereby enhancing GB mobility [51], a mechanism also invoked for other oxides [25,27,28,34–38,40].

In the present study, however, enhanced grain growth is observed near the anode (the positive electrode)

under a steady-state current density of ~40–41 mA/mm$^2$ and higher, following a flash event triggered by an initial electric field of $E$ = 48.50 V/mm. This observation reveals a previously unrecognized mode of anode-side enhanced grain growth in $SrTiO_3$. This study uncovers current-driven elemental redistribution and incomplete redox reactions at junctions where ionic currents are converted into electronic currents, leading to a local reduced region with excess oxygen vacancies and the formation of fast-moving, Sr-rich and reduced GBs. These findings elucidate new mechanisms of defect-mediated microstructural evolution under substantial electric currents and provide new insights into field-matter interactions.

## 2 Experimental

Strontium titanate ($SrTiO_3$) powder (99.95% purity, ~100 nm particle size; US Nanomaterials, Houston, TX, USA) was first calcined at 1100 °C for 2 h to remove potential carbonates and then uniaxially pressed into green pellets (13 mm in diameter) at 130 MPa. The green pellets were subsequently sintered in air at 1450 °C for 4 h, achieving a relative density of ~97%. Each sintered pellet was polished and cut into four specimens with approximate dimensions of 3.5 × 3.5 × 1.14 mm$^3$ (and specific specimen dimensions are listed in Table S1 in the Supplementary Material). To ensure good electrical contact, a Pt layer was sputtered onto both faces of each specimen using a Denton Discovery 18 sputter coater (Moorestown, NJ, USA).

The effects of applied DC electric currents on grain growth were investigated in a tube furnace using a homemade $Al_2O_3$ fixture. The fixture held a sandwich structure consisting of a Pt foil/Pt-sputtered specimen/Pt foil between two alumina plates. The Pt foils were connected to a DC power supply via Pt wires, serving as the cathode and anode, respectively (Figure S1). Current and voltage were recorded using a Tektronix DMM 4040 6½-digit digital precision multimeter. In each run, a DC voltage limit of 55 V and a target (maximum) current were preset before isothermal treatment. An initial voltage of 55 V was applied to trigger a flash-like event in ~1 minute, after which the current was controlled at the preset constant value.

In the first set of experiments, microstructural evolution was studied as a function of annealing duration. The specimens were heated inside a furnace to 1000 °C at 10 °C/min. Upon reaching 1000 °C, the DC power supply was switched on with an initial voltage of 55 V and a preset current limit of 0.5 A, and maintained for 1, 30, 90, 150, or 210 min (at a constant current of 0.5 A after the occurrence of a flash-like event in ~1 minute). Then, both the power supply and furnace were turned off. Because the specimens experienced additional Joule heating under the applied electric current, the actual sample temperatures ($T_S$) were estimated using a blackbody radiation model [31].

The second set of experiments examined the effect of different current limits. An initial voltage limit of 55 V and current limits of 0.25, 0.375, and 2 A were set, with corresponding annealing durations of 210, 210, and 30 min, respectively. The specimen annealed at 2 A was limited to 30 min because of its much faster grain growth.

Notably, microstructural evolution occurred under constant-current conditions, in contrast to prior experiments conducted under negligible currents [46,75,76]. These experiments were therefore designed to specifically investigate the effects of the applied electric current on microstructural evolution.

Specimens were mounted in epoxy, and their cross sections were fine-polished. Microstructures and elemental distributions were examined using an FEI Apreo LoVac scanning electron microscope (SEM) equipped with electron backscatter diffraction (EBSD) and energy-dispersive X-ray spectroscopy (EDS) detectors. EBSD data were post-processed using the MATLAB-based MTEX toolbox [79] for inverse pole figure (IPF) reconstruction and grain size and shape analysis. A custom MTEX-based MATLAB script was also developed to extract data for grain-growth kinetics analysis.

Transmission electron microscopy (TEM) lamellae were prepared using a Thermo Scientific Scios dual-beam focused ion beam (FIB) and SEM system. Atomic-scale GB investigations were performed using an aberration-corrected scanning TEM (AC-STEM; JEOL Grand ARM, JEM-300CF, JEOL, Japan) operated at 300 kV. A direct-electron-detection system (Gatan K2 IS/Summit) was used for electron energy loss spectroscopy (EELS) in counting mode, while a GIF Quantum spectrometer was employed to enhance the information obtained from EELS spectrum-imaging (SI) datasets. EELS SI was acquired in single-EELS mode over an energy range of 400–2400 eV to cover the Ti $L_{2,3}$ edges (~456 eV), O K edge (~532 eV), and Sr $L_{2,3}$ edges (~1940 eV). EELS SI data were processed using the Gatan Microscopy Suite to obtain high-resolution elemental maps and chemical information, with the backgrounds of the core-loss edges removed by power-law fitting.

## 3 Results and discussion

### *3.1 Evolution of electric fields and currents during annealing*

A voltage-to-current control mode was implemented in our experiments, and all specimens exhibited similar voltage-current behavior over time. As shown in Figure 1, an initial constant voltage of 55 V (corresponding to an average electric field of ~48.5 V/mm) was applied to pristine $SrTiO_3$ after the furnace reached 1000 °C. Within approximately 1 min of incubation time, the current initially increased slowly and then surged abruptly upon reaching ~0.2–0.3 A, signaling the onset of a “flash” event with a sudden increase in specimen conductivity, which may be considered as an electrical (dielectric) breakdown coupled with a possible thermal runaway [12,80,81]. When the current reached the preset limit of 0.5 A (corresponding to an average current density of 40.76 mA/mm²), the system switched to constant-current mode and was maintained for different annealing times. Concurrently, the voltage underwent a sudden drop after the flash event and then stabilized at ~14.4–19.6 V (average electric field of ~14.6 V/mm). This voltage drop is attributed to a sudden increase in electrical conductivity to a nominal value of ~2.84 S/m (noting the actual conductivity is not uniform across the specimen thickness, which have hole, ionic, and electron conducting regions, as discussed in Section 4.2). In the steady state, the estimated average specimen temperature increased from 1000 °C to ~1300 °C owing to Joule heating [82], which also contribute to increased specimen conductivity.

The physical origin of the flash event warrants further discussion. When an electric field is first applied, the specimen is expected to be predominantly an insulator or poor ionic conductor via oxygen vacancies at 1000 °C (with blocking Pt electrodes). The rapid increase in current above ~0.2 A can be attributed to a transition to a $p-i-n$ structure with a thin ionic region ($i$-region), thereby reducing the overall specimen resistance, in addition to the temperature effect due to Joule heating. Specifically, defect reactions via $O_O^\times \rightarrow V_O^{\bullet\bullet} + 2e' + \frac{1}{2}O_2 \uparrow$ on the cathode side and via $\frac{1}{2}O_2 + V_O^{\bullet\bullet} \rightarrow O_O^\times + 2h^\bullet$ on the anode side can increase the electron concentration near the cathode and the hole concentration near the anode, respectively [25,28,36,41], thereby forming a $p-i-n$ sandwich structure with a shrinking ionic ($i$) region. These defect reactions can also generate and annihilate oxygen vacancies at the $p-i$ and $i-n$ junctions, thereby sustaining ionic conduction in the ionic region. A thinner ionic region reduces the overall specimen resistance, thereby increasing Joule heating and specimen temperature under constant-voltage conditions during the initial stage. The resulting increase in effective specimen conductivity further amplifies this process through a positive feedback loop, analogous to thermal runaway in flash sintering [12,80,81]. At higher specimen temperatures resulting from Joule heating, Sr vacancies may also become mobile and contribute to ionic conduction in the $i$-region, as well as redox reactions at the junctions. The conduction and associated defect reactions under steady-state, constant-current conditions are discussed further in Section 4.2 (and illustrated in Figure 9 subsequently).

### *3.2 Electric current-driven microstructural evolution*

An unusual grain-growth behavior was observed in $SrTiO_3$ under substantial applied electric currents. As shown in Figure 2(b), elongated grains formed near the anode and were well aligned along the electric-current direction, progressively increasing in size with annealing time. A closer examination of the EBSD map of the specimen after 30 min of annealing under an electric current (Figures 2(b) and 3(a)) revealed that these abnormal grains nucleate in a region near, but not directly at, the anode/$SrTiO_3$ interface, as discussed later. In contrast, no obvious grain growth occurred in the center or near the cathode. These elongated abnormal grains grew near the anode under a substantial electric current density (~40-41 $mA/mm^2$ in this case), in contrast to the previously reported cathode-side enhanced grain growth under virtually current-free conditions, which proceeded isotropically and resulted in uniformly coarsened, equiaxed grains [83].

The grain-size distributions across the specimens (Figure 2(c)) clearly demonstrate that grain growth occurs only near the anode and progressively extends toward the cathode with increasing annealing time, while grain growth elsewhere remains largely dormant. The grain aspect ratios further indicate that these abnormal grains are elongated, with the highest aspect ratio observed at ~90 min.

In summary, these findings indicate that the grains in $SrTiO_3$ can rapidly grow near the anode or the positive electrode under substantial applied electric currents, preferentially along the electric-field direction. This current-induced abnormal grain growth is opposite to previous reports, which consistently observed faster grain growth near the cathode or the negative electrode in $SrTiO_3$ at lower or nominally no electric currents [26,46,51,75,76].

*3.3 Grain growth kinetics for the electric current-driven abnormal grains*

To better understand this growth behavior, the elongated grains were extracted for inverse pole figure (IPF) analysis (Figure 3(a)), and their equivalent diameter and aspect ratio were quantified and mapped in Figure 3(b)–(e). The reconstructed IPFs and grain size (equivalent diameter) and shape maps for the entire cross-sections are provided in Supplementary Figure S2.

With increasing annealing time, the growth region widened along the electric-field direction, increasing by approximately threefold from 30 to 90 min. Grain size and shape evolved differently (Figure 3(b) and (c)). The mean grain size increased approximately linearly with annealing time, from 4.18 ± 2.51 μm at 30 min to 13.93 ± 20.62 μm at 210 min (Figure 3(d)). In contrast, the mean aspect ratio peaked at 2.97 ± 1.87 at 90 min, with a maximum of 16.63, before decreasing to 2.21 ± 0.97 at 210 min (Figure 3(e)). In contrast, the center and cathode regions exhibited negligible changes in mean grain size (~3.34 ± 1.44 μm) and aspect ratio (~1.65 ± 0.47), comparable to those of pristine $SrTiO_3$ (3.13 ± 1.31 μm and 1.62 ± 1.93 respectively). Thus, at 210 min, the mean grain size and aspect ratio of the abnormal grains were approximately 4.5 and 1.8 times those of the normal (dormant) grains, respectively. The observed sharp spatial transition in grain size suggests a transition region across which the oxygen potential drops steeply as a large electric current flows from the anode to the cathode [27,36].

To determine the grain growth kinetics, the grain size $d$ at time $t$ was fitted to a power-law relationship:

$$d^n - d_0^n = kt \tag{1}$$

where $d_0$ is the initial grain size, $n$ is the grain growth exponent reflecting the dominant rate-controlling mechanism [84,85], and $k$ is a temperature-dependent kinetic constant. For the average equivalent diameter of the abnormal grains, a linear relationship ($n$ = 1) provided the best fit, with a slope of 0.05329 (Figure 3(f)). In comparison, $n$ = 2 corresponds to classical normal grain growth governed by capillary-driven GB migration [84,86,87], whereas $n > 2$ is typically associated with reduced GB mobility caused by solute drag, defect interactions, or pinning effects, resulting in slower, diffusion-limited growth.

In prior studies of $SrTiO_3$ without applied electric fields, parabolic grain-growth kinetics has generally been observed and attributed to curvature-driven GB migration ($n$ = 2 and $k = 2\alpha\gamma M$, where $\gamma$ is the average GB energy, $M$ the GB mobility, and $\alpha$ a geometrical factor [87]). A larger $k$ generally indicates higher GB mobility, which depends on temperature, Sr/Ti ratio, oxygen partial pressure, dopants, and GB structure. The growth exponent can exceed 2 when diffusional drag limits GB motion, resulting in decreasing effective GB mobility with increasing grain size, as may occur in the presence of slowly diffusing species such as Sr vacancies or acceptors [51,73,74,87–89].

A grain growth exponent of $n$ = 1 indicates anomalous grain growth kinetics, suggesting that grain growth is governed by mechanisms beyond conventional capillary-driven GB migration. One possible mechanism is vacancy-controlled GB motion, in which GB migration is facilitated by vacancy absorption or generation [84,90,91]. The possible mechanism of vacancy–GB interactions will be discussed further in Section 4.7. Notably, linear grain growth with an exponent of $n$ = 1 has also been reported for anode-side enhanced grain growth in ZnO under an applied electric current [45].

### *3.4 Electric current-driven elemental redistribution*

Near-anode elemental redistribution was observed (Figure 4(a) and Supplementary Figure S3). The EDS maps reveal belt-like regions enriched in Sr but depleted in Ti and O. The Sr-rich belt broadened from an estimated width of ~2.7 µm at 30 min to ~25 µm at 210 min, and migrated toward the cathode with increasing annealing time. Figure 4(b) further reveals that the Sr-rich belt coincides with the grain-size transition interface, corresponding to the front of abnormal grain growth. The abnormal grains grow back into the Ti-rich (Sr-deficient) region, as indicated by the local Sr/Ti ratio, extending toward the anode while simultaneously expanding toward the cathode with the migrating Sr-rich belt. This Sr/Ti redistribution occurs only near the anode (Figure 4(c)). These observations suggest that Sr ions migrate away from the anode and accumulate in the Sr-rich belt, leaving a Ti-rich region with a minimum Sr/Ti ratio of 0.91 after 30 min. This compositional change can be attributed to the formation of substantial Sr vacancies within 30 min, which migrate toward the anode under the electric field.

### *3.5 Grain boundary (GB) characterization*

#### *3.5.1 A general GB in pristine $SrTiO_3$*

A general grain boundary (denoted as "$gb_p$") from pristine $SrTiO_3$ (without an electric field/current) was analyzed by AC-STEM with EELS as a reference. Figure 5(a) shows the HAADF image of $gb_p$ together with EELS elemental maps (at. %) of Sr, Ti, and O. The $gb_p$ core is slightly Sr-rich but Ti- and O-deficient, as also seen in Figure 5(b). In $SrTiO_3$, Ti cations sit at the center of the oxygen octahedra, and the strongly covalent Ti-O bonding splits the $L_2/L_3$ states into the $e_g$ and $t_{2g}$ orbitals [92], which is reflected in the fine structure of the Ti-$L_{2,3}$ edges in EELS [93]. Accordingly, Figure 5(c) reveals four sharp peaks from the Ti-$L_{2,3}$ splitting in respective EELS from the GB core and four nearby regions. The fine structure of the O-K edge shows no clear difference between the GB and bulk region. Together, these are the characteristic $Ti^{4+}$ ions in pristine $SrTiO_3$ [30,94].

EELS can also probe variations in Ti oxidation state [30,58,94–96]. When the Ti valence is lowered by oxygen-vacancy formation, i.e., upon reduction of $SrTiO_3$, distortion of the $TiO_6$ octahedra perturbs the Ti crystal field, broadening the peaks and weakening the splitting of the Ti-$L_{2,3}$ edges. This produces changes in the Ti-$L_{2,3}$ edge intensity that can be used to quantify the reduction in the Ti oxidation state.

*3.5.2 Electric-current effects on GBs compositions in three regions*

The specimen annealed under the electric current for 30 min exhibits three characteristic regions: anode-side normal grains (NG+), abnormal grains adjacent to the Sr-rich belt (AG), and normal grains near the cathode (NG−) (Figure 6(a) and (b)). Notably, both the NG+ and AG regions are located within the Ti-rich bulk region. The effects of electric current on GB composition and Ti valence in these three regions were examined in detail using AC-STEM with EELS. In each instance, two randomly selected GBs were extracted via FIB milling and characterized to show consistency.

In the NG+ region (Figure 6(c)), both randomly selected GBs ($gb1_n^+$ and $gb2_n^+$) are O-enriched, but their Sr and Ti distributions show differences. $gb1_n^+$ is Sr-rich but Ti-poor, whereas $gb2_n^+$ is slightly Ti-rich with a Sr deficiency. Note that the bulk region surrounding $gb2_n^+$ is also Sr-deficient. Towards the electric-field direction, the AG-region GBs ($gb1_{ab}$ and $gb2_{ab}$) in Figure 6(d) are consistently O- and Ti-depleted but Sr-rich, while the bulk region across $gb2_{ab}$ is O-deficient. Compared with $gb_p$ in pristine $SrTiO_3$, the elemental distributions across these GBs are qualitatively similar, but fast-moving $gb1_{ab}$ and $gb2_{ab}$ exhibit stronger Sr segregation and O/Ti depletion. Another key difference is that the abnormal grains (AG) region is highly reduced, as discussed in Section 3.5.3. In the NG+ region, both GBs in the NG- region ($gb1_n^-$ and $gb2_n^-$) are O-enriched (Figure 6(e)). The HAADF images in Figure 6 further reveal largely ordered GB structures in all three regions, without amorphous-like intergranular films observed in some other ceramic GBs [97].

This atomic-scale compositional analysis suggests that the Sr-rich, O/Ti-depleted GBs can enhance grain growth in a Ti-rich bulk region, in contrast to the O-rich GBs associated with normal grains in the NG+ and NG- regions. Prior work has shown that local GB nonstoichiometry, ranging from Ti-rich to neutral and Sr-rich, can occur in both Ti- and Sr-rich bulk compositions of $SrTiO_3$ [62,63,87]. Ti-rich GBs retard GB mobility through solute drag, whereas Sr-rich or neutral GBs exhibit higher mobility, which was attributed to higher concentrations of Ti and O vacancies at the GBs that facilitate mass transport and lower the GB migration barrier [98,99]. This provides a plausible explanation for why Sr-rich GBs with Ti and O depletion may promote abnormal grain growth in $SrTiO_3$ [62,63]. The underlying mechanisms are further discussed and summarized in Section 4.7. It should also be noted that the estimated average sample temperature (~1307 °C in this specimen) is below the eutectic temperatures on both the SrO-rich (1600 °C) and $TiO_2$-rich (1440 °C) sides [100]. Therefore, GB liquid-phase wetting is not expected and can be ruled out as a cause of the observed enhanced grain growth.

### *3.5.3 EELS analysis of Ti valences in GBs in different regions*

STEM-EELS of the Ti-$L_{2,3}$ and O-K edges were acquired from the selected GBs and their neighboring bulk regions to investigate the Ti oxidation state and local oxygen deficiency (Supplementary Figure S4). Although the O-K edge fine structure can reflect O-O ordering and thus oxygen-vacancy formation [101], its weak and noisy signal makes it less sensitive than the Ti-$L_{2,3}$ edges [94]. Therefore, the analysis focuses on the Ti-$L_{2,3}$ peak intensity.

Figure 7(a) shows the Ti-$L_{2,3}$ EELS acquired from each selected GB and the reference $gb_p$. The $t_{2g}$-$e_g$ splitting is weakened to varying degrees at selected GBs, with the peak positions shifting slightly toward lower energy. In particular, the Ti-$L_{2,3}$ spectrum at $gb2_{ab}$ exhibits the characteristic two-peak feature associated with $Ti^{3+}$. These spectral changes indicate disruption of the Ti-O coordination environment due to oxygen-vacancy formation and the associated reduction of $Ti^{4+}$ [93,96,102].

To quantify changes in Ti oxidation states, the *r* value is used to evaluate the relative intensities of the $L_3$-$e_g$ and $t_{2g}$ peaks, defined as $r \equiv \frac{(A-C)}{(B-C)}$, where *A* and *B* are the intensities of the $L_3$-$e_g$ and $t_{2g}$ edges, respectively, and *C* is the intensity of the valley between them, as marked in Figure 7(a). A higher *r* value generally indicates a lower Ti valence [30,94]. The peak-ratio distribution in Figure 7(b) provides the average *r* values for the GBs and their adjacent bulk regions in the pristine and annealed specimens. The *r* value of $gb_p$ (2.55 ± 0.25) is slightly higher than that of its bulk regions (2.12 ± 0.26), while the bulk *r* values in the NG+, AG, and NG− regions remain relatively low, ranging from 1.83 ± 0.22 to 2.63 ± 0.19.

Previous studies suggest that $r$ = 2–2.2 corresponds to negligible Ti reduction associated with subtle oxygen deficiency [30], although *r* can vary with sample-preparation conditions. For example, a minimum *r* of 2 was reported for single-crystal $SrTiO_3$ [30], whereas $r$ = 2.5 was reported for a conventionally sintered polycrystalline sample without Ti reduction [94]. Based on the reported relationship between *r* and oxygen deficiency (δ) in the $SrTiO_3$ lattice, $r$ = 2.1–3.0 corresponds to a maximum δ of 0.07 for a DC-flash-sintered sample [30]. In contrast, *r* values of 5.065 and 10.238 correspond to substantially higher oxygen deficiencies of approximately δ = 0.12 and 0.21, respectively, in oxygen-deficient $SrTiO_{3-\delta}$ films [94,103]. In the present study, $r > 2.63$ is taken to indicate partial $Ti^{4+} \rightarrow Ti^{3+}$ reduction, with an increasing $Ti^{3+}$ fraction at higher $r$ values, particularly for $r > 5$ [30,94].

Along the electric-field direction, the GB $r$ values span a broad range in the NG+ region (2.76 ± 0.04 to 5.66 ± 3.04), reach their highest values in the AG region (5.91 ± 1.16, with $r > 10$ at $gb2_{ab}$), and decrease to a narrower, lower range in the NG− region (3.02 ± 0.21 to 4.13 ± 0.38). The $r$ values >10 for $gb2_{ab}$ and its bulk region are plotted off-scale because the $L_3$-$e_g/t_{2g}$ splitting could not be unambiguously identified (Figure 7(a) and Supplementary Figure S4(b)). In summary, these EELS results indicate that the GBs in the NG regions are susceptible to moderate degree of Ti reduction (more reduced in comparison with the bulk phase), while the fast-moving GBs in the AG region exhibit pronounced $Ti^{4+} \rightarrow Ti^{3+}$ reduction with $r > 5$.

Grain growth in $SrTiO_3$ is well known to accelerate under reducing atmospheres [70–72,78]. Low oxygen partial pressure, corresponding to a high oxygen-vacancy concentration, reduces the space charge and the accumulation of cationic defects (e.g., Sr vacancies) at GBs. The resulting decrease in solute drag to GBs and thereby promotes grain growth [26,51,61,65,66,71,75,78,104]. Conversely, oxidized $SrTiO_3$ develops a pronounced space charge due to Sr-vacancy segregation at GBs, which retards GB mobility and suppresses grain growth [61,64]. In studies of field-induced grain growth in $SrTiO_3$ with blocking electrodes (virtually no currents), an applied electric field drives oxygen vacancies toward the negative electrode, where their accumulation promotes cathode-side grain growth [26,46,75,105].

Overall, these results show that Sr-rich, O-depleted, and highly reduced GBs in the Ti-rich bulk exhibit remarkably high mobility, driving rapid grain growth. The underlying mechanisms may include reduced space-charge effects and solute drag due to suppressed Sr-vacancy segregation in the presence of supersaturated O vacancies in the locally reduced region, reduced GB migration barriers associated with reduction, and enhanced cation transport through Sr- and O-vacancy-assisted diffusion in the Ti-rich bulk region. Related mechanisms are further discussed in Section 4 (and various mechanisms to enhance grain growth are assembled in Section 4.7). The electric-current-driven growth of aligned and elongated abnormal grains also enables the anode-side grain-growth kinetics to follow a linear law with an exponent of $n = 1$.

### *3.6 Effects of the electric current density on microstructural evolution in $SrTiO_3$*

Current-induced microstructural evolution in $SrTiO_3$ was further examined at different current limits using the same voltage-to-current protocol. Specifically, 55 V was applied to trigger the flash event after the furnace reached 1000 °C, followed by application of different preset current limits.

At a low current limit ($I_{limit}$ = 0.25 A), no pronounced abnormal grain growth was observed at the anode even after 210 min, while only weak grain growth occurred near the cathode (Figure 8(a) and Supplementary Figure S5). When $I_{limit}$ exceeded 0.375 A, enhanced (abnormal) grain growth shifted to the anode side (Figure 8(b)–8(d)). Further increasing the current limit promoted anode-side grain growth and expanded the abnormal-growth region; at $I_{limit}$ = 2 A, grain growth was strongly enhanced within only 30 min.

Notably, for $I_{limit} \geq 0.375$ A, the electric field stabilized at ~14.65 ± 1.05 V/mm, whereas for $I_{limit}$ = 0.25 A, it increased to ~41.75 V/mm. This counterintuitive observation is likely associated with the flash-like event, which can substantially alter the generation and transport of charged defects and charge carriers, increasing specimen conductivity through electrical (or dielectric) breakdown. When the current limit is below ~0.2–0.3 A, the flash-like event either does not occur or does not fully develop, limiting the availability of charge carriers needed to sustain the prescribed current. Consequently, the specimen exhibits higher resistivity and a higher electric field, and different grain growth and microstructural evolution behavior.

Although pronounced grain growth was absent at 0.25 A, EDS and EBSD analyses revealed a Sr-rich belt near the cathode rather than the anode (Supplementary Figure S5). Because the Sr-rich belt has been identified as the origin of enhanced grain growth at higher currents (Sections 3.2–3.4), its formation on the cathode side suggests that lower currents may favor cathode-side enhanced grain growth, consistent with previous reports [46,75]. Indeed, this specimen under a lower current showed slight enhanced grain growth with some elongated grains near the cathode (Supplementary Figure S5). The absence of pronounced grain growth under the present conditions is likely attributable to the limited annealing duration and relatively low temperature. In contrast, previous studies reporting cathode-side enhanced grain growth employed substantially longer annealing times (>10 h) in blocked-electrode experiments or higher temperatures (1350–1550 °C) during flash sintering [46,75].

Consequently, these observations demonstrate that microstructural evolution in $SrTiO_3$ depends on the applied electrical loading conditions, with distinct behaviors emerging at high and low current densities and under different experimental conditions. This finding helps reconcile the anode-side and cathode-side enhanced grain growth observed in the present and previous studies [46,75].

# 4 Mechanisms

## *4.1 Defect chemistry*

The voltage application creates two conditions that underline the phenomena observed. First, it establishes an oxygen partial pressure ($P_{O_2}$) gradient within the specimen, with higher $P_{O_2}$ near the anode that decreases toward the cathode. Second, it raises the sample temperature through Joule heating. In undoped $SrTiO_3$, it is generally accepted that the intrinsic O and Sr vacancies ($V_O^{\bullet\bullet}$ and $V_{Sr}''$) can arise predominantly from partial Schottky disorder [106–108], described in the Kröger-Vink notation by

$$\mathrm{Sr_{Sr}^{\times} + O_O^{\times} \leftrightarrow V_{Sr}'' + V_O^{\bullet\bullet} + SrO_{R/P}}\ . \tag{2}$$

Here, $\mathrm{SrO_{R/P}}$ denotes the Ruddlesden-Popper (RP) phase $Sr_{n+1}Ti_nO_{3n+1}$ or $SrO(SrTiO_3)_n$, which consists of rock-salt SrO layers interleaved with $SrTiO_3$ blocks [100,109]. Above ~1225 °C, Sr vacancies become mobile, and the cation kinetics become sufficiently fast to allow near-equilibrium redistribution of the cation sublattice on a reasonable timescale [110]. Because the current surge following the flash event raises the estimated average sample temperature to ~1300 °C (~1307 °C in the 30-min specimen selected for detailed characterization and analysis, as shown in Figures 6 and 7), Sr vacancies become mobile and can actively participate in defect reactions and migration.

The defect chemistry model of Moos and Härdtl [57,111] does not provide equilibrium defect concentrations at high $P_{O_2}$. By combining this model with the partial Schottky disorder reaction, we analyze the distribution of dominant charged defects in $SrTiO_3$ as a function of $P_{O_2}$, as illustrated in the Brouwer diagram in Figure 9(a). We further extend the model to higher $P_{O_2}$ (from $10^0$ to $10^{10}$ bar) and an elevated temperature of 1307 °C to represent the conditions relevant to our experiments.

The defect chemistry model at 1307 °C is developed from the electroneutrality condition:

$$n + 2\,[\mathrm{V_{Sr}''}] = 2\,[\mathrm{V_O^{\bullet\bullet}}] + p, \tag{3}$$

where $n \equiv [e']$ and $p \equiv [h^\bullet]$ are the electron and hole concentrations. The dominant defect reactions, including oxygen incorporation and release, band-to-band transfer, and partial Schottky disorder, together with their mass-action expressions and reaction constants, are summarized in Supplementary Table S2. The reaction constants $K_{oxi}$ , $K_{red}$ , $K_i$ and $K_s$ are each expressed as functions of sample temperature (Supplementary Table S2). All values were adopted from previous studies based on the model of Moos and Härdtl [57,111]. Under the three electroneutrality conditions listed in Supplementary Table S3, the equilibrium concentrations of electrons ($n$), holes ($p$), O vacancies ($[\mathrm{V_O^{\bullet\bullet}}]$), and Sr vacancies ($[\mathrm{V_{Sr}''}]$) at 1307 °C were calculated as functions of $P_{O_2}$ over the range of $10^{10}$ to $10^{-30}$ bar (Figure 9(a)), corresponding to the possible local $P_{O_2}$ conditions established during the electrical loading in our experiments.

*4.2 Formation of p–i–n regions and defect reactions*

In the Brouwer diagram (Figure 9(a)), different charged defects predominate over three distinct $P_{O_2}$ ranges: holes ($h^\bullet$) at high $P_{O_2}$ (light blue region), ionic defects ($V_{Sr}''$ and $V_O^{\bullet\bullet}$) over a narrow intermediate $P_{O_2}$ range (light red region), and electrons ($e'$) at low $P_{O_2}$ (light purple region). These regions correspond to three distinct conduction regimes: *p*-type, ionic (*i*), and *n*-type. When a constant current flows through the specimen after the flash-like event, defect redistribution causes sequential conduction by holes in the *p*-region (anode, +), ionic species ($\mathrm{V_{Sr}''}$ and $\mathrm{V_O^{\bullet\bullet}}$) in the *i*-region, and electrons in the *n*-region (cathode, −).

Consequently, *p–i* and *i–n* junctions are established under the constant applied electric current in a steady state, with the dominant charge carriers transitioning from holes to ionic conduction at the *p–i* junction and from ionic to electronic conduction at the *i–n* junction. Charge continuity requires the current density to be identical throughout the specimen:

$$j_h = j_i = j_e \equiv j \tag{4}$$

In the *i*-region, $V_O^{\bullet\bullet}$ and $V_{Sr}''$ are assumed to be the dominant mobile ionic defects, which contribute jointly to the total ionic current density. The applied electric field drives $V_{Sr}''$ toward the anode and $V_O^{\bullet\bullet}$ toward the cathode. Under the constant-current condition, the total current density is the sum of the individual ionic current contributions:

$$j = j_i = j_{V_O^{\bullet\bullet}} + j_{V_{Sr}''} = 2F(J_{V_O^{\bullet\bullet}} - J_{V_{Sr}''}), \tag{5}$$

where $j$ is the current density in the specimen, $F$ is the Faraday constant, and $J_{V_O^{\bullet\bullet}}$ and $J_{V_{Sr}''}$ denote the $V_O^{\bullet\bullet}$ and $V_{Sr}''$ fluxes in the *i*-region, respectively. The minus sign reflects the opposite charges of $V_O^{\bullet\bullet}$ and $V_{Sr}''$.

As shown in Figure 9(c), to maintain mass and charge continuity, the ionic current associated with Sr vacancies must be converted to holes at the *p–i* junction via the following defect (reduction) reaction:

$$V_{Sr}'' + 2h^{\bullet} + SrO_{R/P} \rightarrow Sr_{Sr}^{\times} + \frac{1}{2}O_2 \uparrow \tag{6a}$$

and to electrons at the *i–n* junction via the following defect (oxidation) reaction:

$$Sr_{Sr}^{\times} + \frac{1}{2}O_2 + 2e' \rightarrow V_{Sr}'' + SrO_{R/P}\downarrow \tag{6b}$$

Equations (6a) and (6b) provide a mechanism for moving the Sr-rich belt (RP phase) toward the cathode, which will be discussed further in §4.4.

Likewise, Figure 9(d) shows that the ionic current associated with O vacancies must be converted to holes at the *p–i* junction via the following defect (reduction) reaction:

$$O_O^{\times} + 2h^{\bullet} \rightarrow V_O^{\bullet\bullet} + \frac{1}{2}O_2 \uparrow \tag{7a}$$

and to electrons at the *i–n* junction via the following defect (oxidation) reaction:

$$\frac{1}{2}O_2 + V_O^{\bullet\bullet} + 2e' \rightarrow O_O^{\times} \tag{7b}$$

Based on Equation (5), we can define the fraction of the ionic current carried by Sr vacancies in the *i*-region as:

$$\lambda \equiv \frac{j_{V_{Sr}''}}{j}, \tag{9}$$

which varies from $\lambda_{p-i}$ at the *p–i* junction (Figure 9(c)) to $\lambda_{i-n}$ at the *i–n* junction (Figure 9(d)), if additional Sr and O vacancies are created in the *i*-region based on Equation (2).

Here, the relative contributions of the defect reactions described by Equations (6a) and (6b) at the *p–i* junction are given by the ratio of $\lambda_{p-i}\colon(1-\lambda_{p-i})$. Likewise, the relative contributions of the defect reactions described by Equations (7a) and (7b) at the *i–n* junction are given by the ratio of $\lambda_{i-n}\colon(1-\lambda_{i-n})$.

*4.3 Formation of the Sr-rich belt and Ti-rich region*

As illustrated in Figure 9(b), O and Sr vacancies can be generated in the *i*-region via the defect reaction given in Equation (2), rewritten below in the direction of O- and Sr-vacancy formation:

$$\mathrm{Sr_{Sr}^{\times}} + \mathrm{O_O^{\times}} \rightarrow \mathrm{V_{Sr}''} + \mathrm{V_O^{\bullet\bullet}} + \mathrm{SrO_{R/P}}\downarrow. \qquad (10)$$

The applied electric field drives negatively charged Sr vacancies $\mathrm{V_{Sr}''}$ toward the anode and positively charged O vacancies $\mathrm{V_O^{\bullet\bullet}}$ toward the cathode, while promoting precipitation of the Sr-rich RP phase, $SrO_{R/P}$. This provides a mechanism for forming the Sr-rich belt (Figure 9(b)). Based on Equation (10), the total precipitation of $SrO_{R/P}$ in the *i*-region should be equal to the total generation of Sr vacancies ($\mathrm{V_{Sr}''}$) in moles, which can be quantified by the following continuity equation for the molar rate per unit area:

$$-\int_{x_{p-i}^{+}}^{x_{i-n}^{-}}\left(\frac{dJ_{V_{Sr}''}}{dx}\right)dx = \frac{j}{2F}\cdot(\lambda_{i-n}-\lambda_{p-i}). \qquad (11)$$

In addition, Figure 9(c) and Equations (6a) and (6b) illustrate that $SrO_{R/P}$ (and an equal amount of Sr vacancies) can also be created at the *i–n* junction (at the rate of $\lambda_{i-n}\cdot j/(2F)$) and annihilated at the *p–i* junction (at the rate of $\lambda_{p-i}\cdot j/(2F)$) if the interfacial reactions are complete. Thus, Equation (11) indicates that there is no net $SrO_{R/P}$ precipitation when all interfacial reactions are complete, i.e., when all ionic currents are converted to electronic currents.

However, the defect reaction at the *p-i* junction (Equation 6(a)) is likely incomplete due to kinetic limitations, as its occurrence requires all $SrO_{R/P}$ to encounter Sr vacancies and electrons to be fully dissolved. If Equation 6(a) at the *n-i* junction is incomplete, there will be net $SrO_{R/P}$ precipitation, leading to the formation and growth of the Sr-rich region, as observed experimentally.

In addition, the incomplete interfacial reaction Equation 6(a) (or incomplete consumption of $\mathrm{V_{Sr}''}$ at the *p-i* junction) implies that some Sr vacancies can penetrate (drift by the electric field) to the *n*-region, creating a Ti-rich region between the Sr-rich belt and the anode, which was also observed experimentally.

*4.4 Redox-modulated migration of the Sr-rich belt*

Figure 9(c) illustrates a redox-modulated mechanism for the migration of the Sr-rich belt. The Sr-rich

RP phase can dissolve/annihilate at the *p–i* junction via Equation 6(a) and precipitate at the *i–n* junction via Equation 6(b), in addition to the possible precipitation within the *i*-region via Equation (10). The combination of these defect reactions provides a mechanism for effectively moving the Sr-rich belt along the field direction. This redox-modulated mechanism for the migration of the Sr-rich belt explains the experimental observation shown in Figure 2(b).

*4.5 Local reduction*

Figure 9(d) illustrates another set of interfacial reactions involving oxygen redox, in which holes are converted to oxygen vacancies ($V_O^{\bullet\bullet}$) at the *p–i* junction via Equation 7(a), and oxygen vacancies are converted to electrons at the *i–n* junction via Equation 7(b). The oxygen vacancies migrate toward the *i–n* junction, where they are annihilated through oxygen incorporation, which consumes $O_2$ gas and electrons, as described by Equation 7(b). However, the $O_2$ supply is likely limited by slow kinetics, including diffusion of $O_2$ and interfacial reaction kinetics. Consequently, the oxidation reaction at the *i–n* junction is likely incomplete, leading to an excess of oxygen vacancies ($V_O^{\bullet\bullet}$) and creating a locally reducing environment. This local reduction will promote grain growth, as discussed in Section 4.7.

*4.6 Pore formation and nucleation of abnormal grains*

This set of reactions illustrated in Figure 9(d) and described by Equations 7(a) and 7(b) is accompanied by oxygen bubbling at the *p–i* junction via Equation 7(a) and by void formation through the condensation of supersaturated oxygen vacancies when the oxidation reaction at the *i–n* junction via Equation 7(b) is incomplete. Such pores have been reported to serve as nucleation sites [25,41]. As shown in Figure 10(b), elongated grains begin to nucleate near the Sr-rich belt after 30 min (with abnormal grains in the Ti-rich region, presumably with the migration of the Sr-rich belt). The magnified images in Figures 10(c) and 10(d) clearly reveal pore formation coinciding with the abnormal grains.

As shown in Figure 10(c), abnormal grains do not nucleate directly at the anode but rather within the *p–i–n* junction region near the anode, presumably promoted by local reduction associated with supersaturated oxygen vacancies. This locally reduced region with supersaturated oxygen vacancies can be considered a "virtual cathode" [112]. After nucleation, these abnormal grains rapidly grow into the Ti-rich region to reach the anode, while also extending toward the cathode along with the migrating Sr-rich belt.

*4.7 The mechanisms of abnormal grain growth*

Under the synergistic effects of cation redistribution, characterized by the formation and migration of the Sr-rich belt and the widening of the Ti-rich region, and the locally reducing environment discussed above, reduced GBs enriched in Sr form, promoting the growth of elongated abnormal grains in the Ti-rich bulk region. These abnormal grains have fast-moving, Sr-rich and O-depleted GBs (Figure 6(d)), where $Ti^{4+}$ is partially reduced to $Ti^{3+}$ (Figure 7). The underlying mechanisms responsible for the enhanced mobility of these GBs are elaborated and summarized below.

First, the local reduction discussed in Section 4.5 leads to reduced GBs with O depletion and Ti reduction. As discussed in Section 3.6, grain growth in $SrTiO_3$ is known to accelerate under reducing atmospheres [70–72,78], which has been attributed to reduced space charges, Sr-vacancy segregation and the associated reduction in solute drag [61,64]. Prior studies of field-induced grain growth in $SrTiO_3$ with blocked electrodes have suggested that the local accumulation of oxygen vacancies can promote grain growth (and in this case, the applied electric field drives oxygen vacancies toward the negative electrode, promoting cathode-side grain growth) [26,46,75]. In the present case, GBs are generally more reduced than the bulk, as indicated by EELS (Figure 7), suggesting preferential segregation of oxygen vacancies to GBs. With a supersaturation of oxygen vacancies (discussed in Section 4.5), substantial oxygen vacancies can segregate to GBs, further increasing GB oxygen deficiency as evident in Figure 9(d) (noting vacancies are not well-defined at GBs upon relaxation). Notably, the fast-moving GBs for the abnormal grains exhibit a higher degree of $Ti^{4+}$-to-$Ti^{3+}$ reduction (Figure 7), suggesting a possible valence-state-dependent mechanism for enhanced GB mobility [113]. Another prior study also suggested that reduction can induce a GB structural transition that triggers abnormal grain growth in $Bi_2O_3$-doped ZnO [114]. Our EELS results show that $Ti^{4+}$ in these GBs is partially reduced to $Ti^{3+}$ (Figure 7), which may similarly enhance the mobility of the reduced GBs, even though a GB structural transition is not evident in the present case.

Second, the Sr-rich belt can lead to enhanced Sr segregation at GBs, as confirmed experimentally (Figure 6(d)). As discussed in Section 3.5.2, prior studies have suggested that Sr-rich GBs exhibit higher mobility, because the higher concentrations of Ti and O deficiencies at these GBs may facilitate mass transport and lower the GB migration barrier, whereas Ti-rich GBs exhibit lower mobility due to solute drag associated with Sr vacancies [98,99].

Third, even if the fast-moving, Sr-rich, reduced GBs nucleate within the Sr-rich belt near the $i-n$ junction that is most reduced, the abnormal grains rapidly grow into the widening Ti-rich bulk region toward the anode as the Sr-rich belt rapidly migrates toward the cathode, while the GBs remain Sr-rich or Ti-depleted even within the Ti-rich bulk region, as observed experimentally. Ti-rich $SrTiO_3$ ($Sr/Ti < 1$) has been reported to promote grain growth. From a defect-chemistry perspective, a Ti-rich bulk composition can introduce bulk Sr and O vacancies, facilitating cation transport through vacancy-assisted mechanisms to further enhance grain growth [62,63,68,73,74,87,115–118].

In addition, the elongation and anisotropic growth of abnormal grains may be driven by vacancy flux (along with vacancy generation and annihilation), analogous to the vacancy–GB interaction models [90,91], in which GB motion is strongly coupled to vacancy redistribution. In these models, GBs migrate toward regions of higher vacancy concentrations, while their migration facilitates the creation and absorption of vacancies, thereby enhancing GB diffusion and migration.

In summary, the coupled redistribution of Sr and O vacancies, local reduction, and the formation of a moving Sr-rich belt and a widening Ti-rich region create favorable conditions for the formation of Sr-rich, O-depleted, and highly reduced GBs that grow into the Ti-rich bulk region. These GBs exhibit high mobility (based on the three possible mechanisms summarized above) and interact with vacancy fluxes, leading to the growth of aligned, elongated abnormal grains along the field/flux direction, with linear kinetics distinct from classical parabolic grain growth.

### 4.8 *Summary of microstructural evolution under applied electric currents*

Figure 11 summarizes the proposed mechanisms of microstructural evolution under applied electric currents. Following the initial flash event (an electrical/dielectric breakdown possibly coupled with a thermal runaway [9–12]), *p–i–n* regions form. To sustain a constant current density after the flash, charged-defect transport is maintained by redox defect reactions at the junctions, leading to oxygen bubbling and RP-phase dissolution at the *p–i* junction, and oxygen incorporation and RP-phase precipitation at the *i–n* junction. Sr and O vacancies can also be generated in the ionic region and driven by the applied electric field toward opposite electrodes.

Incomplete annihilation of Sr vacancies at the *p–i* junction allows some Sr vacancies to drift into the *p*-region by the electric field, creating a Ti-rich region near the anode and an adjacent Sr-rich belt that grows and migrates toward the cathode. Meanwhile, incomplete oxidation of O vacancies at the *i–n* junction leads to the formation of a locally reduced region. Together, these processes create favorable conditions for the formation of Sr-rich, O-depleted, and reduced ($Ti^{4+} \rightarrow Ti^{3+}$) GBs, triggering abnormal grain growth. The abnormal grains then grow into the Ti-rich bulk region, where the bulk defect population further promotes grain growth. These abnormal grains also extend toward the cathode, following the migrating Sr-rich belt. Vacancy–GB interactions drive grain elongation along the vacancy flux direction, resulting in linear growth kinetics that are distinct from classical parabolic grain growth.

## 5 Conclusions

In this work, polycrystalline $SrTiO_3$ was employed as a model material to investigate microstructural evolution under substantial electric currents. In contrast to prior reports of cathode-side-enhanced grain growth under current-blocking conditions, a high current density (e.g., ~40-41 mA/mm$^2$) induces abnormal grain growth near the anode.

Following an initial flash-like event after ~1 minutes, the specimen conductivity increases, and the system switches to constant-current mode, leading to the formation of *p–i–n* regions in which electronic and ionic currents are interconverted through redox reactions at the *p*–*i* and *i*–*n* junctions. Incomplete interfacial redox reactions lead to the formation of a widening Ti-rich region near the anode, a migrating Sr-rich belt, and a locally reduced region with supersaturated oxygen vacancies. Together, these features create favorable conditions for the formation of fast-moving, Sr-rich, and reduced GBs, driving the linear-in-time growth of aligned, elongated abnormal grains along the field/flux direction, in contrast to classical parabolic grain growth.

These findings reveal previously unrecognized mechanisms of microstructural evolution in $SrTiO_3$ under substantial electric currents, providing new insights into field–matter interactions in oxide ceramics. More broadly, field–matter interactions represent a fundamental materials science problem with broad technological implications, including understanding microstructural stability in electronic devices and electrochemical cells under electric currents and developing novel materials processing approaches to tailor microstructures.

**Acknowledgments:** This work was supported by the Materials of Extreme Properties program of the U.S. Air Force Office of Scientific Research (AFOSR) under Grant No. FA9550-22-1-0413. We thank our program manager, Dr. Ali Sayir, for his support and advice. We also thank Dr. Toshihiro Aoki for assistance and suggestions on AC-STEM and EELS acquisition and Dr. Shu-Ting Ko for assistance on MTEX analysis.

**Supplementary Materials**

Supplementary material related to this article, including Table S1-S3 and Figures S1-S5, can be found in the online version at doi: xxxxxx.

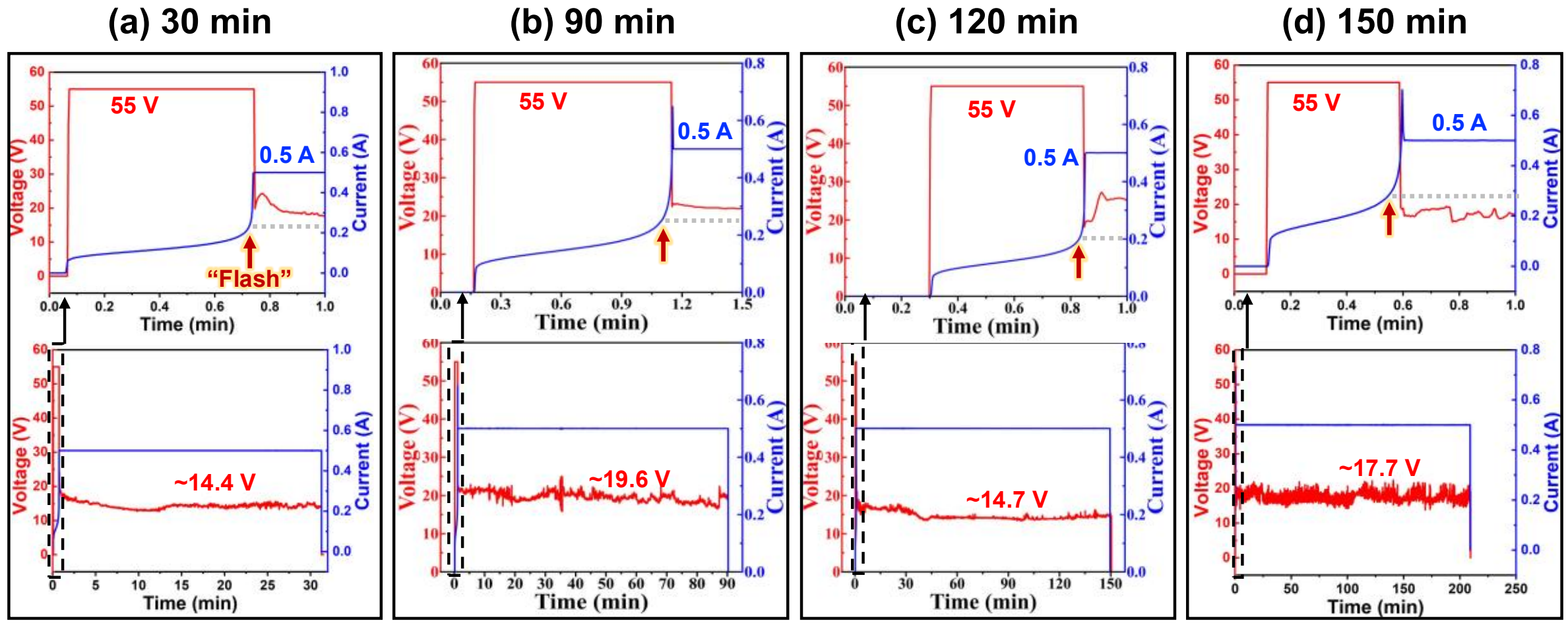

**Figure 1.** Temporal profiles of voltage (*U*) and current (*I*) during isothermal annealing experiments at 1000 °C under applied electrical bias for (a) 30 min, (b) 90 min, (c) 150 min, and (d) 210 min. In all instances, an initial voltage of 55 V was applied in constant-voltage mode, followed by a flash event occurring between ~0.5-1.1 min (as illustrated in the magnified insets of the *U-I* response during the initial minute), after which the system transitioned to a preset maximum current of 0.5 A under constant-current control.

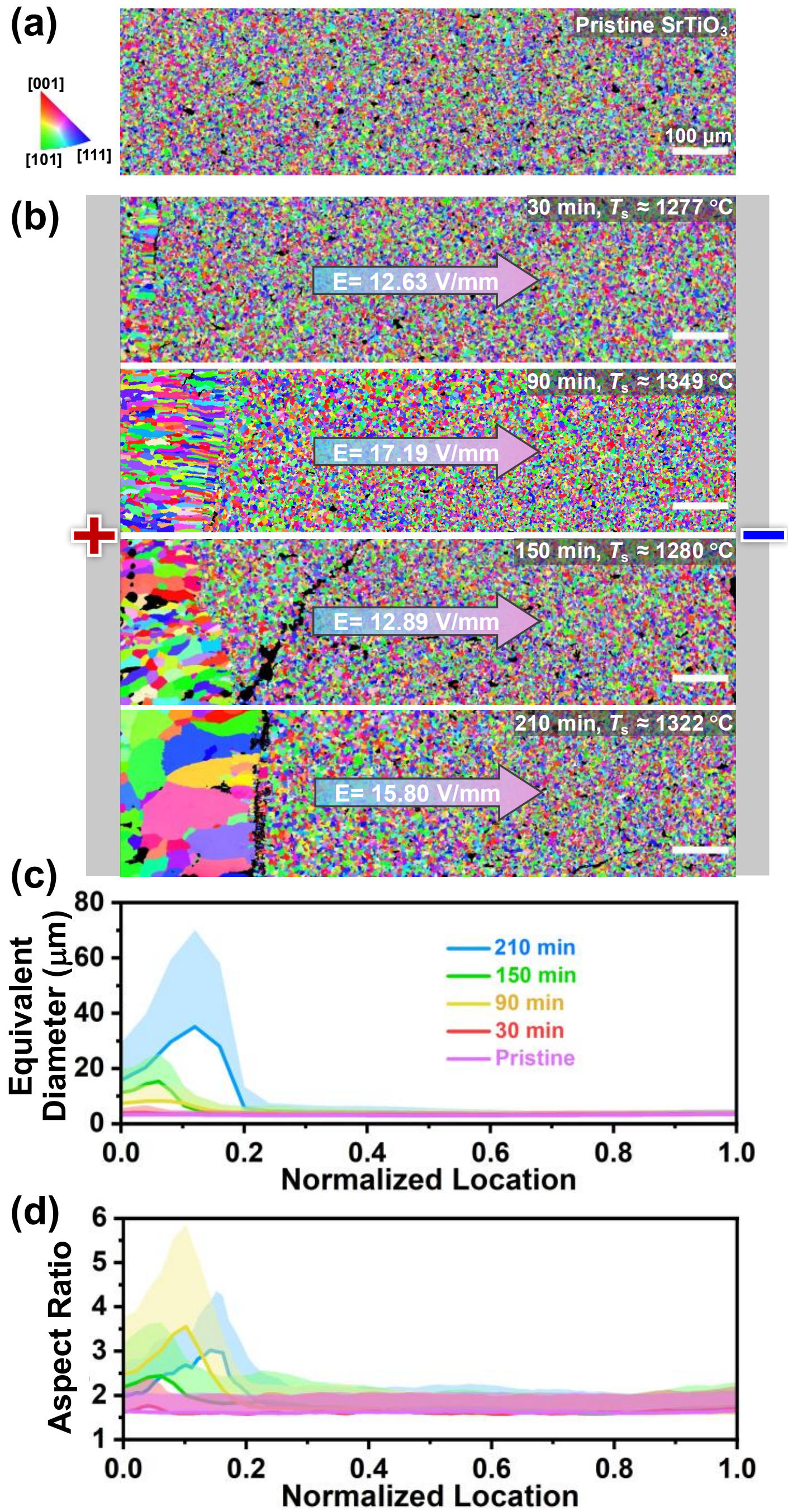

**Figure 2.** Electron backscatter diffraction (EBSD) maps of cross-sections for (a) a pristine specimen and (b) specimens annealed under electrical loads (with a constant current of 0.5 A or ~40-41 mA/mm$^2$ current density after an initial flash event) for 30, 90, 150, and 210 min, respectively. (c) Grain sizes and (d) grain shape factors (grain aspect ratios) as a function of normalized location, where shaded areas denote +1 standard deviation.

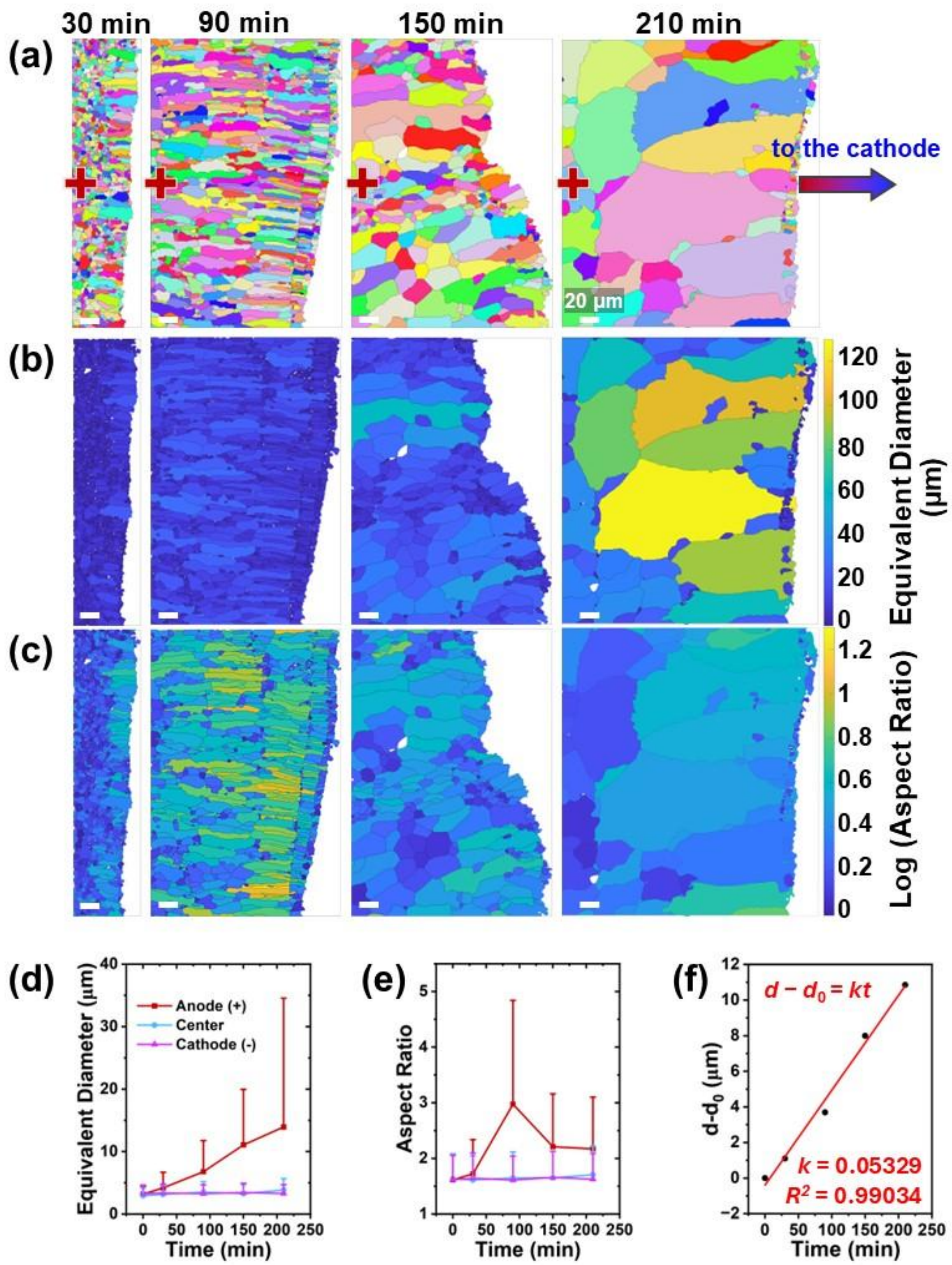

**Figure 3.** (a) Reconstructed inverse pole figures (IPFs) of the abnormal grain growth regions adjacent to the anode. Equivalent diameter maps (b) and aspect ratio maps (c) are displayed on a logarithmic scale. Corresponding IPFs and grain size and shape maps for the complete cross-sections are provided in Supplementary Figure S2. (d) Temporal evolution of the average equivalent diameter and (e) the aspect ratio of abnormal grains at the anode, contrasted with normal grains located in the center and cathode regions of the specimens. (f) Linear regression of the average grain diameter as a function of time for abnormal grains adjacent to the anode.

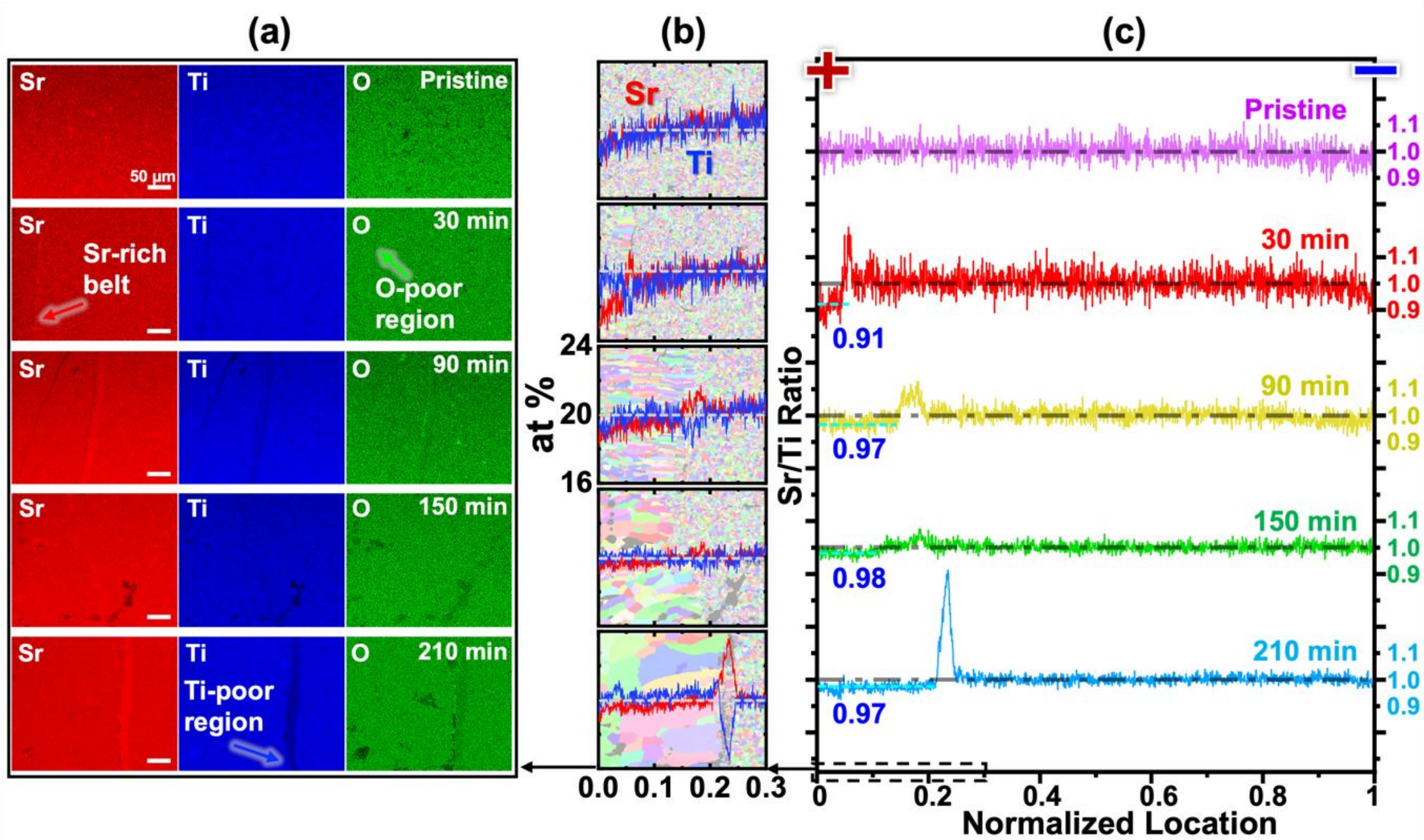

**Figure 4.** (a) Energy-dispersive X-ray spectroscopy (EDS) elemental maps of Sr, Ti, and O acquired from the near-anode regions of various specimens. Full EDS maps for the entire specimens are provided in Supplementary Figure S3. (b) Compositional profiles (at. %) of Sr and Ti as functions of normalized location to the anode, overlaid with the corresponding EBSD maps. (c) Ti ratio as a function of normalized location, derived from the EDS maps presented in Supplementary Figure S3.

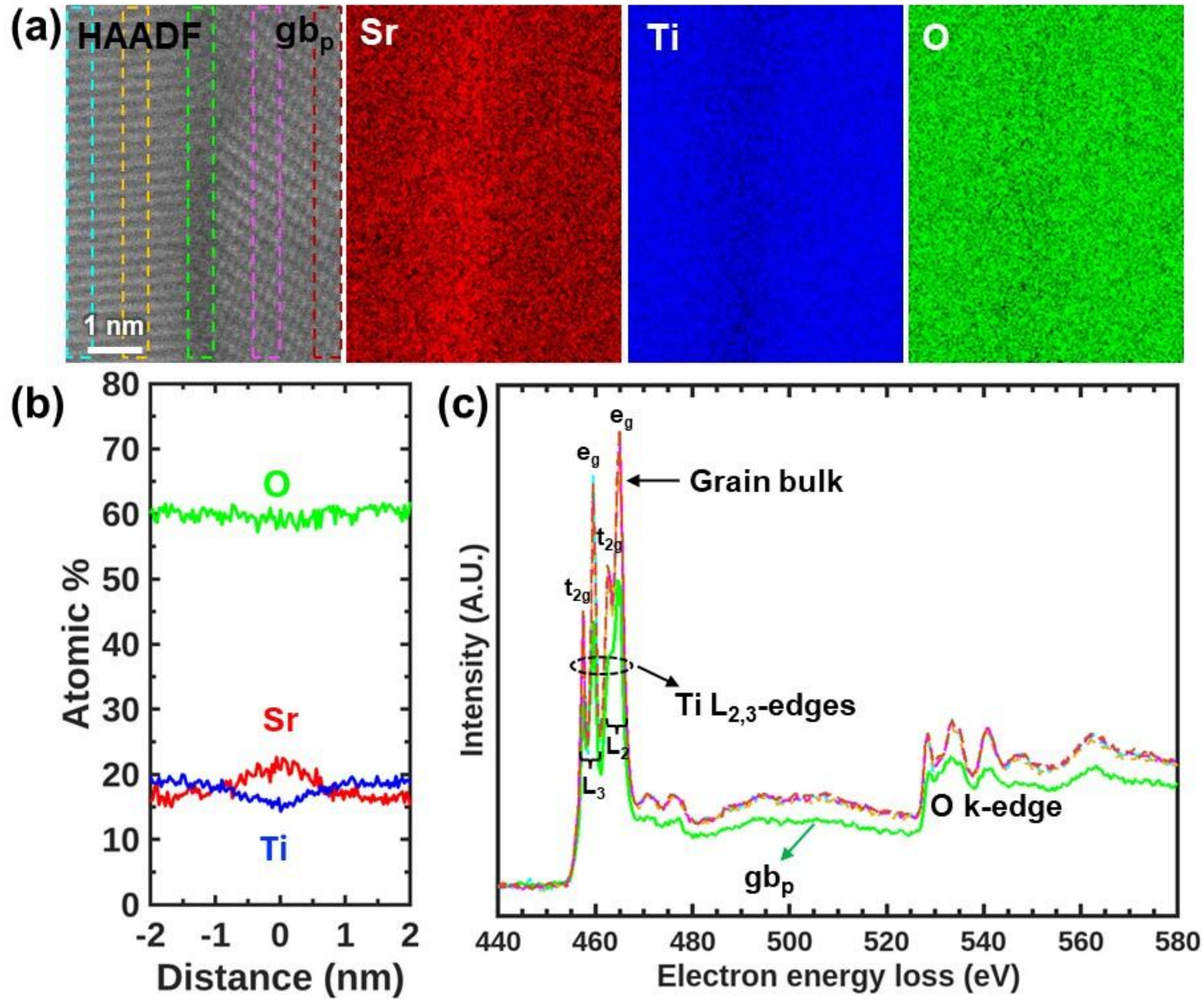

**Figure 5.** (a) Aberration-corrected scanning transmission electron microscopy (AC-STEM) high-angle annular dark-field (HAADF) image of a grain boundary ($gb_p$) in an as-sintered specimen (prior to annealing under electrical bias), accompanied by electron energy-loss spectroscopy (EELS) elemental maps Sr, Ti, and O. (b) One-dimensional (1D) compositional line profiles across the grain boundary. (c) Ti-$L_{2,3}$ edges and O-K edge spectra acquired from five regions indicated by colored dashed boxes in (a), demonstrating the intensity contrast between the grain boundary (light green) and the bulk material (the remaining four locations, which exhibit nearly overlapping spectra).

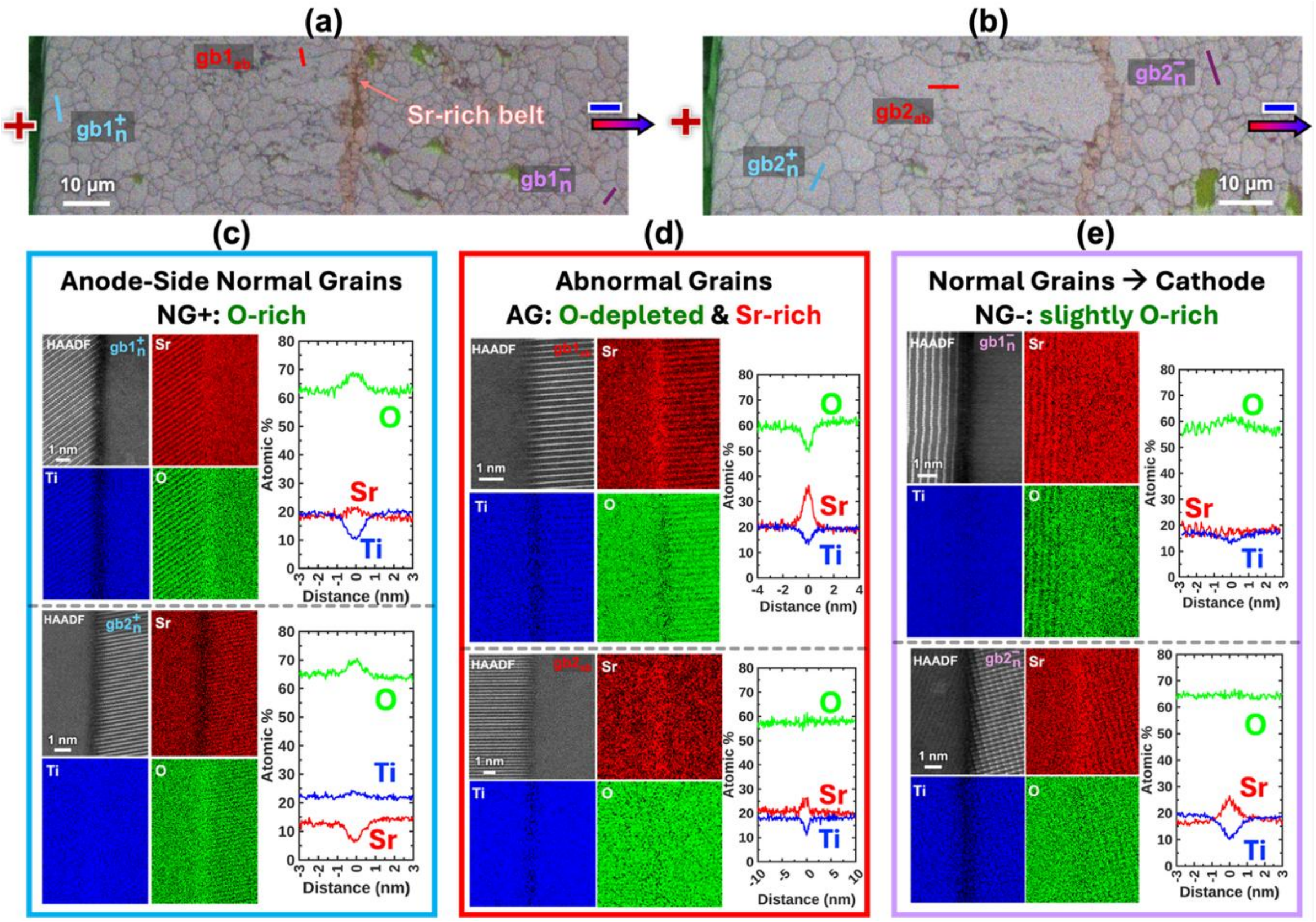

**Figure 6.** (a, b) Band-contrast images overlaid with SEM-EDS maps of the near-anode region of the specimen annealed under an applied current of 0.5 A for 30 min, where the locations of six (3×2) grain boundaries (GBs) extracted via a focused ion beam (FIB) are indicated. AC-STEM HAADF images accompanied by EELS elemental maps of Sr, Ti, and O and 1D compositional line profiles across specific GBs for the locations indicated in (a) and (b): (c) $gb1^{+}_{n}$ and $gb2^{+}_{n}$ for GBs from anode-side normal grains, (d) $gb1_{ab}$ and $gb2_{ab}$ from abnormal grains, and (e) $gb1^{-}_{n}$ and $gb2^{-}_{n}$ from normal grains approaching to the cathode. In each instance, two randomly selected GBs were extracted via FIB and characterized to show consistency.

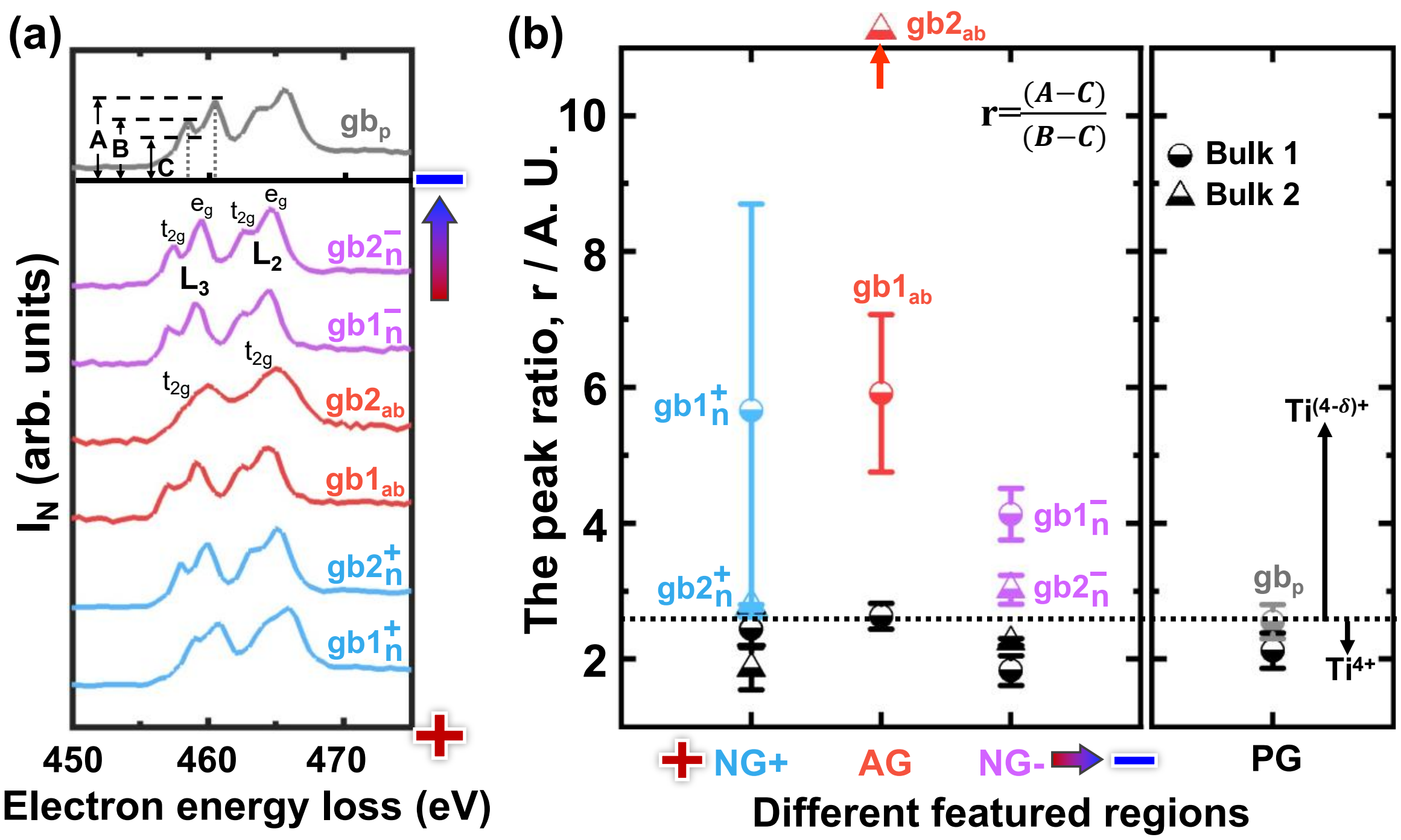

**Figure 7.** (a) Electron energy-loss spectroscopy (EELS) spectra of the Ti-$L_{2,3}$ edges acquired from six (3×2) grain boundaries (GBs) in the specimen annealed under an applied current of 0.5 A for 30 min. (b) Distributions of the peak ratio, $r$ values calculated from the bulk materials and GBs in anode-side normal grains (NG+), abnormal grains (AG), and normal grains approaching the cathode (NG-), as compared to the GBs in pristine grains (PG). The $r$ value is defined as $(A-C)/(B-C)$, where A, B, and C are marked in (a), to characterize the valence of Ti. A high $r$ value implies a high degree of reduction of $Ti^{4+}$ to $Ti^{3+}$ [30,94].

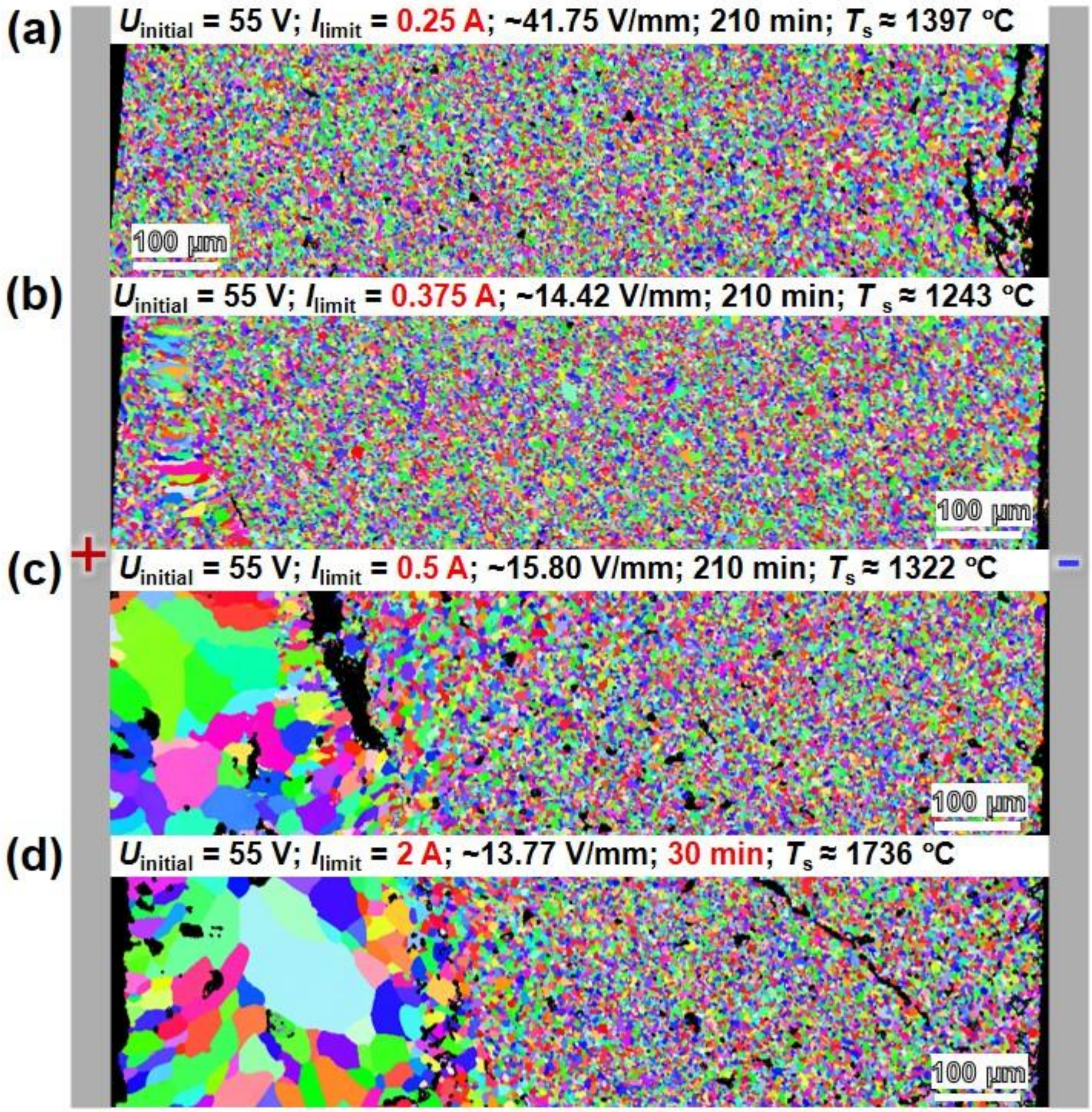

**Figure 8.** Electron backscatter diffraction (EBSD) maps of $SrTiO_3$ specimens annealed under different applied electric currents with varying preset limits of (a) 0.25 A, (b) 0.375 A, (c) 0.5 A, and (d) 2 A. An initial voltage of 55 V was applied in all instances. Annealing durations and estimated specimen temperatures are provided.

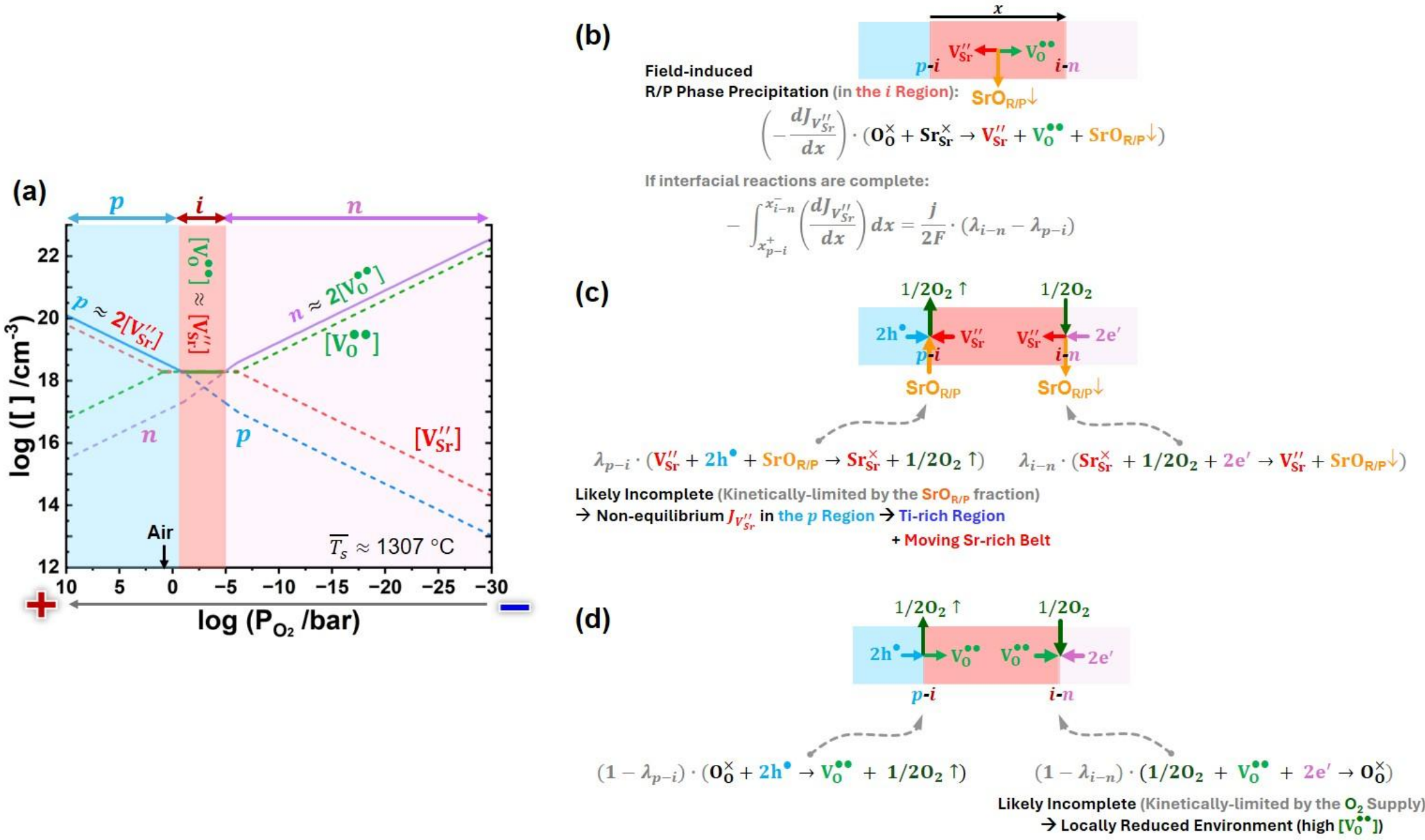

**Figure 9.** (a) A Brouwer diagram calculated using the Moos and Härdtl's model [57,111], showing the defect concentrations in $SrTiO_3$ at an average sample temperature of 1307 °C as a function of oxygen partial pressure ($P_{O_2}$). The dominant defects, electrons ($e'$), holes ($h^{\bullet}$), Sr, and O vacancies ($V_{Sr}''$ and $V_O^{\bullet\bullet}$), are drawn as solid lines, with the electroneutrality conditions indicated in each region. The hole (*p*-type), ionic (*i*), and electron (*n*-type) conduction regions are shaded light blue, red, and purple, respectively. Schematic illustrations of (b) field-induced generation and separation/drift of Sr and O vacancies and RP-phase $SrO_{R/P}$ precipitation in the *i*-region, and the conversion between ionic and electronic currents at the *p*–*i* and *i*–*n* junctions through possible interfacial defect reactions involving (c) $V_{Sr}''$ and (d) $V_O^{\bullet\bullet}$. In (b), the minus divergence of Sr vacancy flux ($-dJ_{V_{Sr}''}/dx$) represents the Sr-vacancy generation rate per unit volume, equivalently the $SrO_{RP}$ precipitation rate per unit volume, in the *i*-region. In panel (c), $\lambda_{p-i}$ and $\lambda_{i-n}$ denote the relative contributions of $V_{Sr}''$ to the total ionic current, or equivalently the fractions of the two interfacial reactions involving $V_{Sr}''$, at the *p*–*i* and *i*–*n* junctions, respectively. Correspondingly, the fractions of the two interfacial reactions involving $V_O^{\bullet\bullet}$ at the *p*–*i* and *i*–*n* junctions shown in (d) are $(1-\lambda_{p-i})$ and $(1-\lambda_{i-n})$, respectively. The incomplete interfacial reaction (reduction of $V_{Sr}''$) at the *p*–*i* junction can lead to the drift of the Sr vacancies ($J_{V_{Sr}''}$) into the *p*-region, which promote the formation and growth of the Sr-rich belt through net RP-phase precipitation, as well as the development of a widening Ti-rich region toward the anode. Similarly, the incomplete reaction (oxidation of $V_O^{\bullet\bullet}$) at the *i*–*n* junction can lead to local reduction and excess O vacancies.

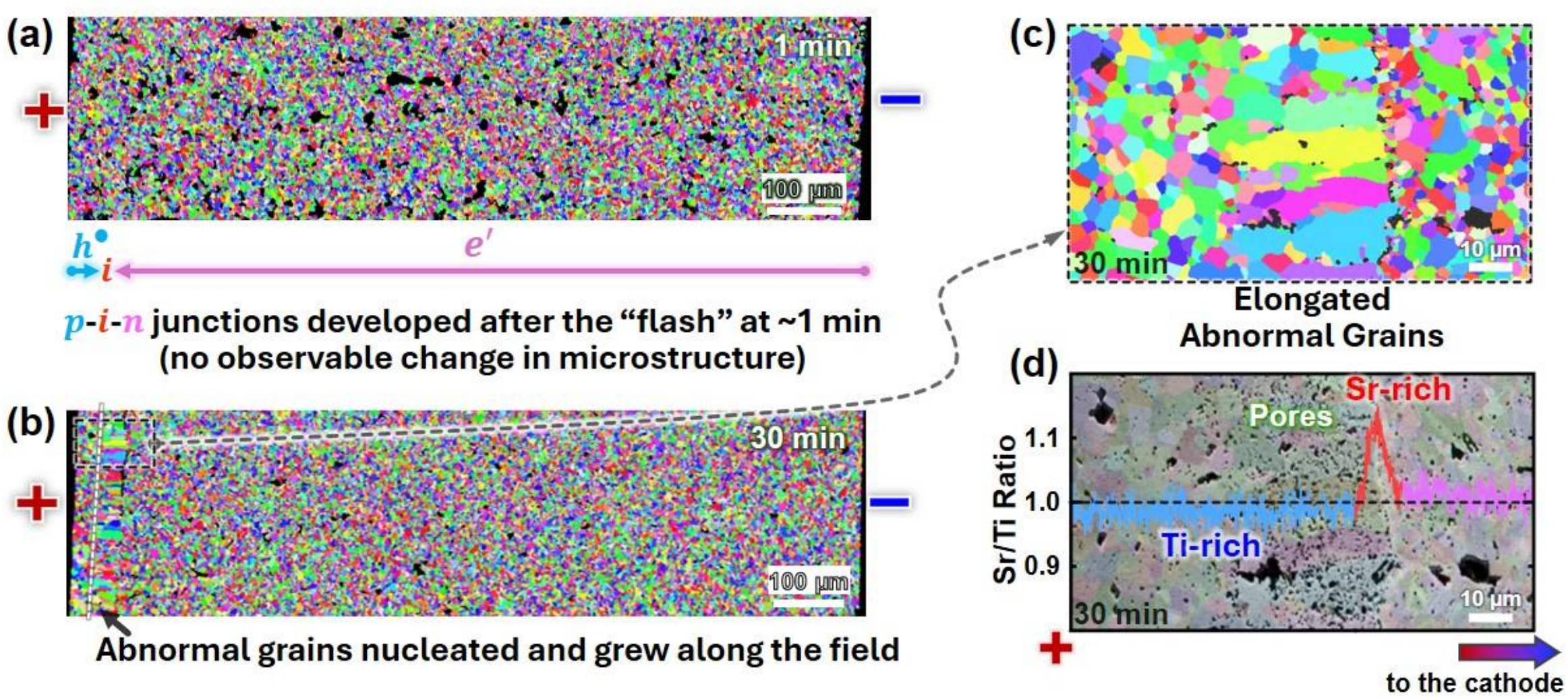

**Figure 10.** Defect and microstructural evolution illustrated via EBSD maps of specimens annealed under applied electrical bias after (a) 1 min (wherein *p–i–n* junctions are presumably formed following the "flash" event that transitioned the system from a constant voltage of 55 V to a pre-set maximum/constant current of 0.5 A) and (b) 30 min (depicting the nucleation and elongated growth of abnormal grains along the field direction). (c) An EBSD map and (d) an FSD image overlaid with the EBSD map of a magnified view of the abnormal grains for the specimen shown in (b). In panel (d), the Sr/Ti ratio profile is overlaid on the images, demonstrating Ti-rich abnormal grains and a Sr-rich belt.

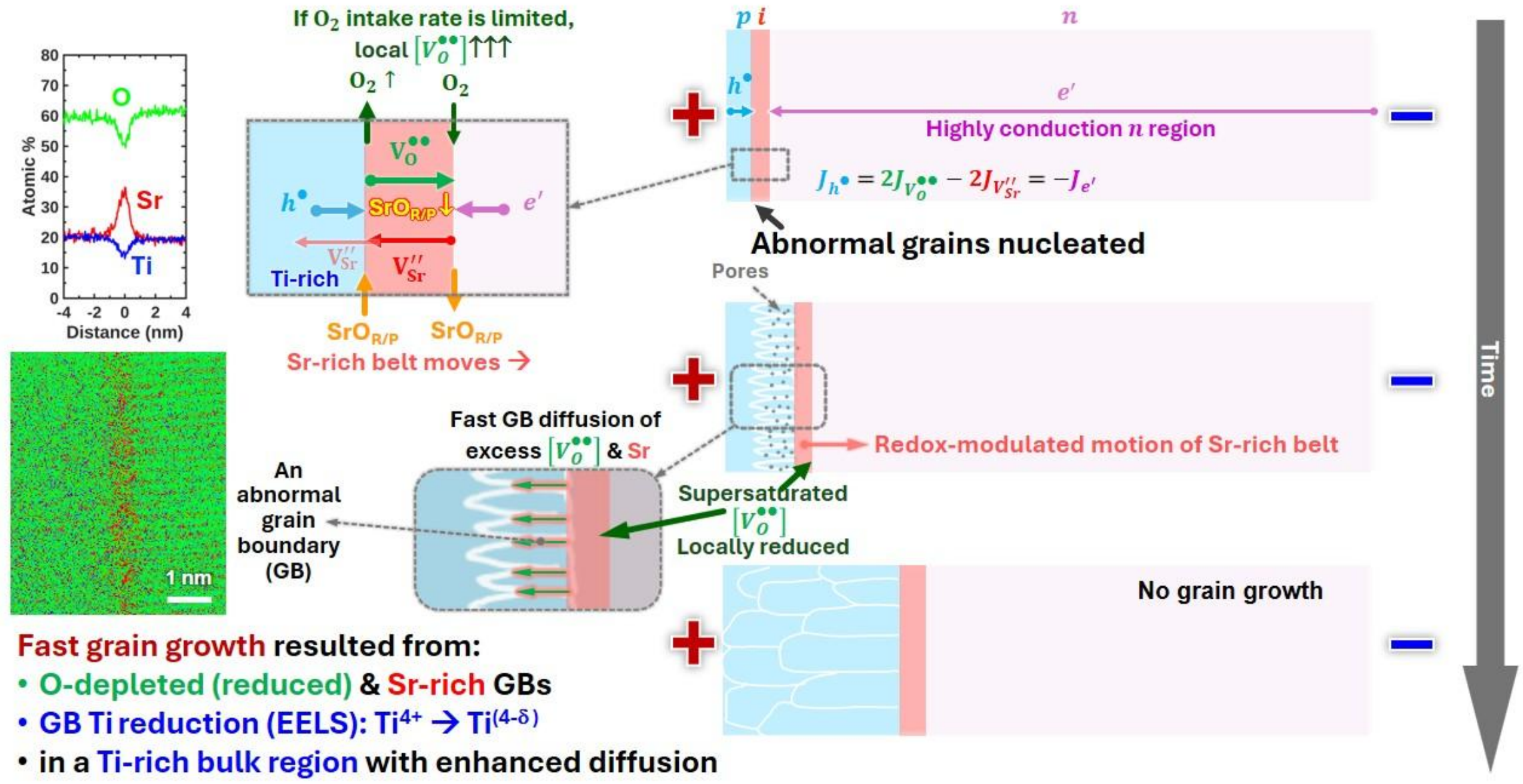

**Figure 11.** Schematic illustration of the proposed mechanisms underlying electric current-driven microstructural evolution in $SrTiO_3$. Following a flash-like event (electrical/dielectric breakdown), hole ($p$-type), ionic ($i$), and electronic ($n$-type) conduction regions form sequentially under a constant electric current. The RP phase ($SrO_{R/P}$) precipitates within the $i$-region and at the $i$–$n$ junction, while a Sr-rich belt and a Ti-rich region can form due to incomplete consumption of the Sr vacancy flux and $SrO_{R/P}$ phase at the $p$–$i$ junction. The transition between electronic and ionic currents drives $O_2$ bubbling at the $p$–$i$ junction. The incomplete oxidation of $V_O^{\bullet\bullet}$ at the $i$–$n$ junction, kinetically limited by the local $O_2$ supply, creates a locally reduced environment with supersaturated oxygen vacancies. The annihilation and generation of $SrO_{R/P}$ at the $p$–$i$ and $i$–$n$ junctions, respectively, provide a mechanism of redox-modulated migration of the Sr-rich belt toward the cathode. Together, these features create favorable conditions for the formation of fast-moving, Sr-rich, O-depleted, and Ti-reduced GBs within the locally reduced region. Facilitated by vacancy–GB interactions, elongated abnormal grains grow along the vacancy flux (or field) direction into the Ti-rich bulk region toward the anode, while simultaneously extending toward the cathode with the migrating Sr-rich belt.